\documentclass{aa}  

\usepackage{graphicx}
\usepackage{natbib}
\bibpunct{(}{)}{;}{a}{}{,} 
\usepackage{txfonts}
\usepackage{multirow}
\usepackage{nicefrac}
\usepackage[rightcaption]{sidecap}

\usepackage[switch, modulo]{lineno}

\begin{document}

\title{Flare-induced changes of the photospheric magnetic field
       in a $\delta$-spot deduced from ground-based observations}


   \author{P. G\"{o}m\"{o}ry\inst{1},
           H. Balthasar\inst{2},
           C. Kuckein\inst{2},
           J. Koza\inst{1},
           A. M. Veronig\inst{3},
           S. J. Gonz\'{a}lez Manrique\inst{2,4},\\
           A. Ku\v{c}era\inst{1},
           P. Schwartz\inst{1},
            \and
           A. Hanslmeier\inst{3}
          }
    \authorrunning{P. G\"{o}m\"{o}ry et al.}
   
    \institute{Astronomical Institute of the Slovak Academy of Sciences,
               05960 Tatransk\'{a} Lomnica, Slovakia\\
               \email{gomory@astro.sk}
          \and
               Leibniz Institut f\"{u}r Astrophysik Potsdam (AIP), An der
               Sternwarte 16, 14482 Potsdam, Germany
          \and
               IGAM-Kanzelh\"{o}he Observatory, Institute of Physics, 
               University of Graz, Universit\"{a}tsplatz 5, 8010 Graz, Austria      
          \and
               Universit\"at Potsdam, Institut f\"ur Physik und Astronomie,
               Karl-Liebknechtstra{\ss}e 24/25, 14476 Potsdam-Golm, Germany       
              }

   \date{Received; accepted:}

 
  \abstract
   {}
   {Changes of the magnetic field and the line-of-sight velocities in the
    photosphere are being reported for an M-class flare that originated
    at a $\delta$-spot belonging to active region NOAA 11865. 
   }
   {High-resolution ground-based near-infrared spectropolarimetric observations
    were acquired simultaneously in two photospheric spectral lines, 
    \ion{Fe}{i} 10783\,\AA\ and \ion{Si}{i} 10786\,\AA, with the Tenerife Infrared
    Polarimeter at the Vacuum Tower Telescope (VTT) in Tenerife on 2013 October 15.
    The observations covered several stages of the M-class flare. Inversions
    of the full-Stokes vector of both lines were carried out and the results
    were put into context using (extreme)-ultraviolet filtergrams from the
    Solar Dynamics Observatory (SDO). 
   }
   {The active region showed high flaring activity during the whole observing
    period. After the M-class flare, the longitudinal magnetic field did not
    show significant changes along the polarity inversion line (PIL). 
    However, an enhancement of the transverse magnetic field of approximately 550\,G
    was found that bridges the PIL and connects umbrae of opposite polarities
    in the $\delta$-spot. At the same time, a newly formed system of loops
    appeared co-spatially in the corona as seen in 171\,\AA\ filtergrams of
    the Atmospheric Imaging Assembly (AIA) on board SDO. However, we cannot
    exclude that the magnetic connection between the umbrae already existed in the
    upper atmosphere before the M-class flare and became visible only later
    when it was filled with hot plasma. The photospheric Doppler velocities show
    a persistent upflow pattern along the PIL without significant changes due to
    the flare.}
   {The increase of the transverse component of the magnetic field after the
    flare together with the newly formed loop system in the corona support
    recent predictions of flare models and flare
    observations.
   }

   \maketitle
%

\section{Introduction} \label{Introduction}

Sunspots of complex magnetic configuration harboring both magnetic polarities
within one penumbra are called $\delta$-spots 
\citep{1960AN....285..271K, 1965AN....288..177K}.            
According to
\cite{1987SoPh..113..267Z},                                  
they can form in three main ways: i) a single structure emerges with reverse
polarity with respect to the Hale-Nicholson rules; ii) satellite dipoles emerge
close to existing spots and the emerging flux region expands, converting a
preceding (in the sense of solar rotation) spot into a following spot (or vice
versa); and iii) a collision between two dipoles may occur so that opposite
polarities are pushed together.

$\delta$-spots are often associated with flares, but not always.
\citet{2012SoPh..281..599T}                                    
observed downflows of 1.5 -- 1.7 km\,s$^{-1}$ in a decaying $\beta\gamma\delta$
active region with a lifetime of 12 hours, and no flares happened during
this phase.
\cite {2014A&A...562L...6B, 2014ASPC..489...39B}               
observed a $\delta$-spot in a quiet phase and found that the magnetic field showed
a smooth transition from the main umbra to that of opposite polarity ($\delta$-umbra).
The $\delta$-umbra developed its own Evershed flow, which stopped at a dividing line
separating the spot into two parts. Next to the $\delta$-umbra, this line falls together
with the polarity inversion line (PIL), but at some distance from the $\delta$-umbra,
this dividing line separates from the PIL. Along the dividing line, the infrared line
\ion{Ca}{ii}\,8542\,\AA{} exhibited a central emission, and in some locations along
this dividing line strong chromospheric up- and downflows of approximately 8\, km\,s$^{-1}$
were seen. A shear flow was detected along the dividing line.
\citet{2014A&A...562L...6B}                                    
explain these results with emerging bipolar flux at the edge of the already existing spot
\citep[`scenario 2' of][]{1987SoPh..113..267Z}.                  
\cite{2014ApJ...789..162C}                                     
studied the magnetic configuration and dynamics along the PIL of a $\delta$-spot
observed with the CRISP instrument at the Swedish Solar Telescope (SST) shortly
after a C4.1 flare was produced in the target region. Their results show upflows
and downflows of around $\pm$3\,km\,s$^{-1}$ in the proximity of the PIL, and also
horizontal motions along the PIL of the order of $\approx$1\,km\,s$^{-1}$.
\begin{figure*}[!t]
    \centering{
    \includegraphics[width=\hsize]{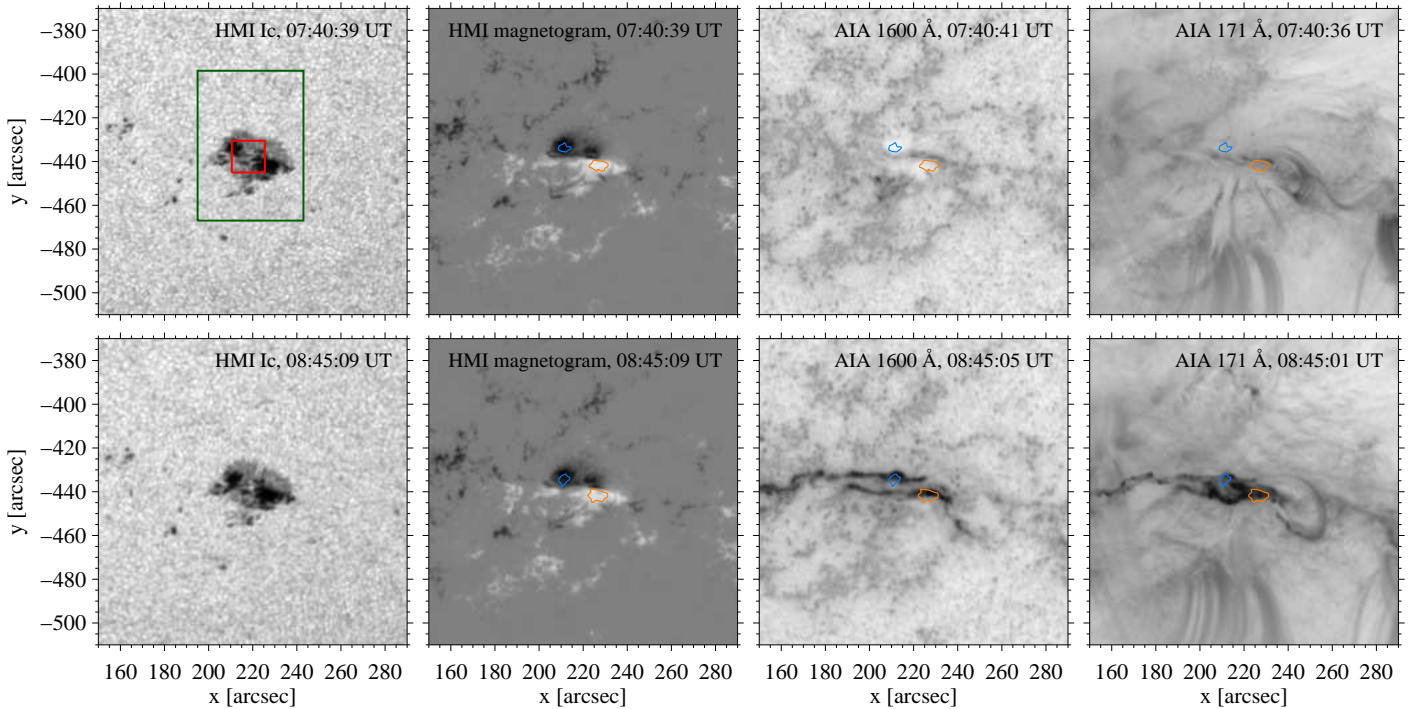}}
     \caption{Overview images showing AR 11865 shortly before {\em (top panels)}
              and during {\em (bottom panels)} observations. The first two panels
              of each row show the HMI continuum intensity and line-of-sight
              magnetic field (values are clipped at $\pm$1500\,G), respectively.
              The last two panels are AIA filtergrams taken in the 1600\,\AA\ and
              171\,\AA\ channels. They are displayed on logarithmic scale and a
              reversed intensity scaling is used. The times listed in the upper
              right corner of each panel correspond to the midpoint of the particular
              HMI or AIA exposure time. The light-blue and orange contours represent
              the umbrae of opposite polarity within the $\delta$-spot. The green
              rectangle outlines the full field-of-view of the TIP-II instrument.
              The red square shows the area displayed in Figs.\,\ref{B_Fe}, \ref{B_Si},
              and \ref{tip_vel}.}
     \label{context}
\end{figure*}

It is often assumed that flares are triggered when shear flows along the PIL
build up magnetic shear or twist. Flow fields in a complex and highly active
sunspot group were investigated by
\citet{2006ApJ...644.1278D}.                                   
They found a shear flow at the PIL of approximately 1\,km\,s$^{-1}$ persisting for five
hours before an X10 flare occurred. However, after the flare these flows were
enhanced and the gradient of the magnetic field changed.
\citet{2009ApJ...690.1820T}                                    
observed a shear flow of 0.6 km\,s$^{-1}$ in a $\delta$-spot between the two
umbrae of opposite polarity. After an X3.4 flare, this flow reduced to
0.3 km\,s$^{-1}$. During flaring activity in a $\delta$-spot,
\citet{2012A&A...547A..34F}                                    
detected Doppler velocities of $\pm$10 km\,s$^{-1}$.
\citet{1994ApJ...425L.113M}                                    
found velocities up to 14\,km\,s$^{-1}$.
\citet{2002ApJ...575.1131L}                                    
reported converging flows towards the PIL in the penumbra of a $\delta$-spot.
However,
\citet{2007SoPh..245..219D}                                    
showed that the magnetic shear was not changed significantly by a photospheric
shear flow.

Flares are associated with a sudden change in the magnetic field topology.
Nevertheless, there are only a few observations which allow us to analyze
topological changes of the magnetic field within a $\delta$-spot which are
related to a flare. Lately, several studies reported changes in the magnetic
field vector during flares. 
\cite{2008ASPC..383..221H}                                     
predicted that the magnetic field vector changes into horizontal fields during
the flare. This is supported by recent findings of
\cite{2012ApJ...745L..17W}                                     
who found a significant increase of the horizontal magnetic field along the PIL 
related to flare activity.
In contrast,
\cite{2015ApJ...799L..25K,2015IAUS..305...73K}      
reported a strong decrease of the magnetic field strength during the flare,
that is, a decrease of both horizontal and vertical components.

Here, we present an analysis of changes of the magnetic field topology of a $\delta$-spot
caused by an M-class flare using a unique ground-based data set acquired with the VTT.
The data consist of three spectro-polarimetric scans of an area covering a $\delta$-spot
during high flaring activity. We focus on variations of the transverse and longitudinal
components of the magnetic field. The spectro-polarimetric findings are discussed in the
context of high-cadence (E)UV imaging by SDO/AIA.

%
\section{Data and data reduction} \label{observations}

Active region (AR) NOAA 11865 was observed on 2013 October 15. The ground-based
observations were performed with the Tenerife Infrared Polarimeter
\citep[TIP-II;][]{2007ASPC..368..611C}                    
at the Vacuum Tower Telescope
\citep[VTT;][]{1998NewAR..42..493V}                       
on Tenerife (Spain). As context data, we also used co-temporal (E)UV filtergrams
provided by the Atmospheric Imaging Assembly
\citep[AIA;][]{2012SoPh..275...17L}                       
and line-of-sight magnetograms, Dopplergrams, and continuum intensity images obtained
by the Helioseismic and Magnetic Imager
\citep[HMI;][]{2012SoPh..275..207S, 2012SoPh..275..229S}. 
Both these instruments are on board the Solar Dynamics Observatory 
\citep[SDO;][]{2012SoPh..275....3P}.                      
The disk-center coordinates of the target were $(x,y) = (220\arcsec,-440\arcsec)$,
and the cosine of the heliocentric angle was $\mu = 0.86$. The global overview of
the observed area shortly before and during the measurements is shown in
Figure\,\ref{context}.

\subsection{VTT data} \label{observations_vtt}
The TIP-II instrument was operated in spectropolarimetric mode and measured 
combinations of the four Stokes parameters along the spectrograph slit. The 
chosen 1\,$\mu$m spectral window comprised two photospheric spectral lines  
which are formed at slightly different photospheric layers and are sensitive
to the magnetic field: \ion{Fe}{i} 10783\,\AA\ and \ion{Si}{i} 10786\,\AA. For
details about these spectral lines we refer to 
\cite{2008A&A...488.1085B}, for example.                                

Three scans (hereafter scans 1, 2, and 3) of 140 steps each with a step size
of 0\farcs35 were taken between 07:45 and 08:09\,UT, 08:10 and 08:33\,UT, and
08:37 and 09:00\,UT. The exposure time for the infrared camera of TIP-II was
250\,ms and the number of accumulations was set to eight. The data calibration
included dark-current subtraction, flat-field correction, and the standard
polarimetric calibration including a residual cross-talk correction
\citep{1999ASPC..184....3C,2003SPIE.4843...55C, 2002A&A...381..668S}.  
Data were binned by a factor of two in spectral and spatial directions. Thus,
the resulting pixel size along the slit was 0\farcs36. The observed spectrum
was compared with the near infrared Fourier transform spectrometer atlas
\citep{1993aps..book.....W},                                
and a spectral sampling of 22.17 m\AA\ per pixel was determined. The inversion
procedure requires a normalization of the Stokes profiles to the quiet-Sun
intensity at disk center. Thus, we multiplied the data with a limb darkening
factor of 0.971 that we obtained from Eq.\,2 of
\cite{1977SoPh...52..179P}.                                 

Inversions of the full Stokes profiles recorded by the TIP-II instrument
were computed with the Stokes Inversions based on Response functions code
\citep[SIR;][]{1992ApJ...398..375R}.                        
The two spectral lines were inverted independently with a one-component
model atmosphere. The temperature stratification of this model corresponds
to the Harvard Smithsonian Reference Atmosphere
\citep[HSRA,][]{1971SoPh...18..347G}                        
and covers the range $-4.4<\log \tau<1.0$. The inversions were performed
in three cycles. For the temperature we use one node in the first cycle and
three nodes in the second and third cycles. Magnetic field strength, inclination,
and azimuth as well as the Doppler velocity were treated as height independent 
(one node was used for each cycle). We subtracted a fixed amount of 7.5\%
dispersed stray light from the quiet Sun, and applied a macro turbulence
of 1.75\,km\,s$^{-1}$.
\begin{figure}[!t]
  \includegraphics[width=\columnwidth]{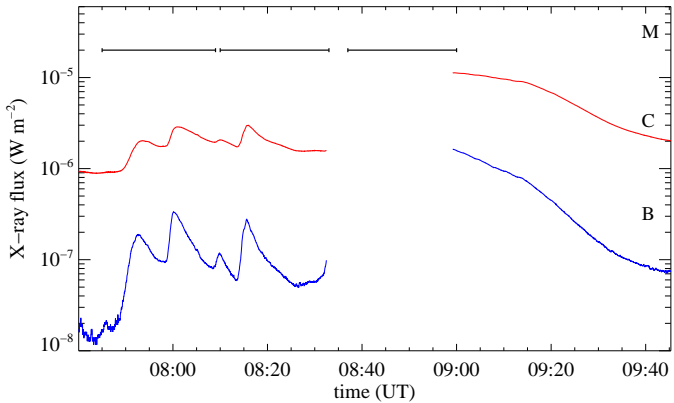}
  \caption{GOES 1--8\,\AA\ (red line) and 0.5--4.0\,\AA\ (blue line) light curves
           showing the temporal evolution of the X-ray flux during ground-based
           observations. We note the gap in the data during the occurrence of the
           strongest event. The three horizontal black lines correspond to the
           time of scans 1, 2, and 3 of the VTT.} 
    \label{fig_goes}
\end{figure}

We tried to solve the magnetic azimuth ambiguity with different methods such as
the azimuth center method
\citep[see][]{2006A&A...449.1169B},                         
and the method of
\citet{2009SoPh..260...83L},                                
but for this complex case we could not find a satisfying solution. Therefore,
we consider only the total magnetic field strength and its longitudinal and
transverse components in the following.

\subsection{SDO data} \label{observations_sdo}
The SDO/AIA instrument provides multiple, high-resolution full-Sun images
(field of view $\sim$1.3\,$R_\odot$). It carries four telescopes and
quasi-simultaneously observes solar transition region and corona at a
12-second cadence in seven different EUV filters, and at a 24-second
cadence in two UV filters. The spatial resolution of the acquired filtergrams
is $\sim$1\farcs5 with a corresponding pixel size of 0\farcs6\,$\times$\,0\farcs6.

The SDO/HMI instrument takes full-disk observations of the Sun at six
wavelength positions distributed at $\pm$34.4\,m\AA, $\pm$103.2\,m\AA, and
$\pm$172\,\AA\ around the center of the photospheric \ion{Fe}{i} 6173.34\,\AA\
absorption line. The acquired filtergrams are combined to form simultaneous
continuum intensity images, line-of-sight magnetograms and Dopplergrams at
a 45-second temporal cadence and with an image scale of 0\farcs5 per pixel.
The longitudinal magnetograms are obtained with a noise level of 10\,G.
\begin{figure}[!t]
    \includegraphics[width=\hsize]{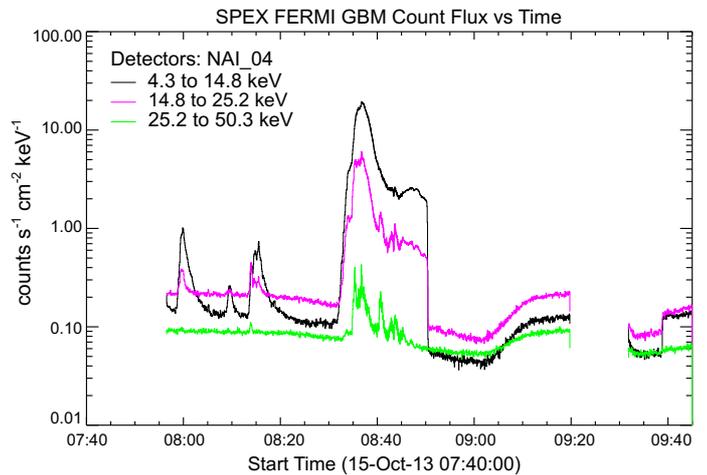}
        \caption{FERMI (4--50\,keV, color-coded lines) X-ray light curves
                covering the same time range as shown in Fig.\,\ref{fig_goes}.
                The displayed data provide complementary information as they fill
                the gap in the GOES measurements.}
        \label{fig_fermi}
\end{figure}

AIA and HMI data were downloaded from the Virtual Solar Observatory (VSO) in
level 1.0 and 1.5 format, respectively. In this format the data from both
instruments have already been partially processed. For AIA data, the pre-processing
includes: bad-pixel removal, despiking and flat-fielding. But the data were not
exposure-time corrected. In case of HMI data, they were also photometrically
corrected. Moreover, images of the physical observables (i.e., continuum images,
magnetograms, and Dopplergrams) were already created from individual filtergrams.
All data were then processed using aia$\_$prep.pro
which is part of the SolarSoft software library implemented within the Interactive
Data Language (IDL). This adjusts all of the different AIA filter images as well as
HMI data products to a common plate scale so that they share the same centers, rotation
angles, and image scales. At the end, all data were spatially derotated to the same
reference frame in order to compensate the differential solar rotation.

As the plasma parameters can evolve quickly in active regions, the observation times
have to be stated very precisely when comparing data from different instruments.
Therefore, we always use the midpoint of the particular exposure (given in UT) if
we refer to individual exposures of AIA and HMI data products.
\begin{figure*}[!t]
  \includegraphics[width=\hsize]{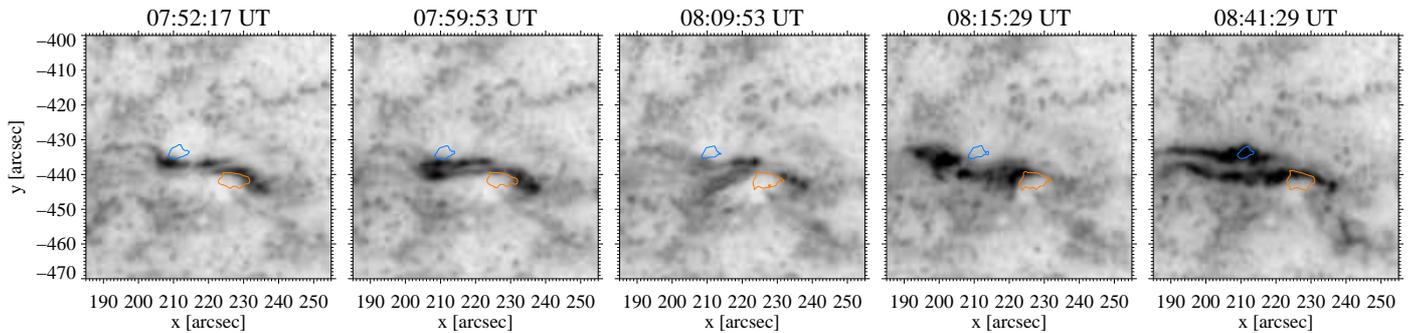}
     \caption{Sequence of AIA 1600\,\AA\ filtergrams showing almost continuous
              flaring activity within the $\delta$-spot of the AR 11865 during
              observations. The logarithmic scale and a reversed intensity scaling
              is used to display the data. The midpoint of the recording time of the
              particular images is presented above each panel. The light-blue
              and orange contours represent the umbrae of negative and positive polarity
              within the spot, respectively. The full time resolution animation of the
              displayed wavelength channel is shown in the online movie.
              } 
     \label{a1600_activity}
\end{figure*}
\begin{figure*}[!t]
  \includegraphics[width=\hsize]{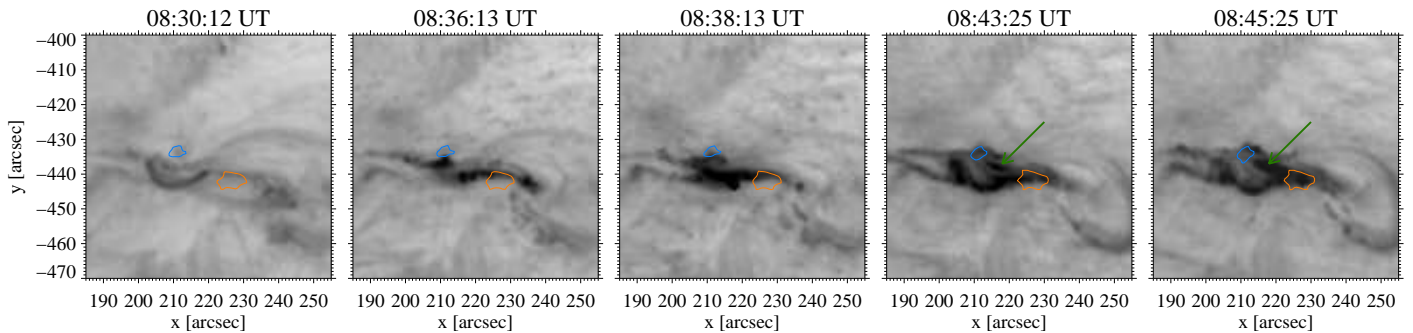}
     \caption{AIA 171\,\AA\ filtergrams at five different times showing temporal
              changes of coronal structures during the observed M-class flare. In particular,
              the images show the temporal evolution of flare loops (marked by the green arrows)
              which connect the umbrae within the $\delta$-spot. This feature was not visible
              before, during the impulsive phase and the peak time of the flare (first three
              panels) but appeared shortly after the flare's X-ray flux reached its maximum (last
              two panels). The logarithmic scale and a reversed intensity scaling is used again.
              The times above each panel correspond to the midpoint of the particular recording
              time. The contours have the same meaning as in Fig.\,\ref{a1600_activity}.
              The full time resolution animation at this wavelength is also available
              as a part of the online movie.
              }  
     \label{a171_loops}
\end{figure*}
\section{Results} \label{Results}

\subsection{Activity in AR 11865}

The NOAA AR 11865 was initially detected on the visible part of the solar disk on
2013 October 9 and was classified as $\beta$-type according to the Hale classification.
However, already on the next day, the $\delta$-configuration of the leading spot
developed. Since that time, the AR 11865 activated and produced C-class flares every
day and also,  exceptionally, M-class events. On the day of our observations (2013 October
15), the active region became highly active and produced eight C- and one M-class flares.

The recorded GOES data (see Fig.\,\ref{fig_goes}) shows that four C-class flares
of slightly different magnitude as well as one M-class flare appeared during our
ground-based observations. There is a gap in the GOES data exactly at the moment
when the strongest flare appeared because the satellite entered the Earth's shadow.
Thus, we cannot identify its real amplitude. The existing data cover only the declining
phase of that flare and indicate that the flare level was at least M1.8. Therefore,
we refer to it as an M-class event in the following text.

Fortunately, the FERMI satellite
\citep{2009ApJ...697.1071A},                                            
captured the temporal evolution of the
X-ray emission during this event. Figure\,\ref{fig_fermi} shows the light curves
in three FERMI energy bands for the same time range as displayed in Fig.\,\ref{fig_goes}.
The time between 07:55\,UT and 08:50\,UT is well covered by FERMI measurements, thus
showing four out of five events detected by GOES (only the first event is missed).
More importantly, the impulsive as well as peak phase of the main M-class event (but
not its decay) is detected. Thus, the combination of the FERMI and GOES measurements
allow us to  also determine precise timing for the strongest event. It started at 08:31\,UT,
peaked at 08:38\,UT (there is also secondary peak visible at 08:48\,UT) and lasted longer
than the ground-based observations. At 08:50\,UT, FERMI also entered Earth's shadow, and
the signals dropped suddenly. Our ground-based observations did not cover the main
impulsive phase of the flare very well. 

The very high activity related to the studied $\delta$-spot during its observations
is demonstrated also in Fig.\,\ref{a1600_activity} (see attached movie). It shows a
series of the AIA 1600\,\AA\ images taken at the moments when the flares indicated by
GOES and FERMI X-ray light curves occurred (cf. Figs.\,\ref{fig_goes} and \ref{fig_fermi}).
The activity is mostly visible along the PIL which separates opposite magnetic polarities
within the $\delta$-spot but some of the stronger C-class eruptions
also show a typical
two-ribbon structure with the ribbons parallel to the PIL.
  
The origin of the largest event is also located in the $\delta$-spot of AR 11865 but in
this case it extensively expands to its surroundings reaching also AR 11864. The AIA
1600\,\AA\ filtergrams show two clearly discernible bright ribbons starting on 08:39\,UT,
which separate in opposite directions. Their roots are at the PIL of the $\delta$-spot.
The two ribbons are seen in all AIA wavelength bands.

Flaring activity related to the $\delta$-spot was also visible in the corona.
For example, the AIA 171\,\AA\ channel, which probes coronal temperatures, exhibited lots 
of flare-related dynamics. The activity occurred mainly along the PIL (i.e., co-spatially
with the flare brightenings observed in the 1600\,\AA\ channel) and only very sporadic and
short-lived magnetic loops bridging the neutral magnetic line were detected during
four C-class events (for details see online movie). Then, a significant new system of hot
loops, which connected the umbrae within the $\delta$-spot, appeared during the strongest flare.

The evolution of this coronal structure is shown in Fig.\,\ref{a171_loops}. The first
three panels of Fig.\,\ref{a171_loops} represent moments slightly before and during the impulsive
phase, and during the peak time of the M-class flare. Here the flare related dynamics started to
be obvious again with increasing time but the brightenings were still localized mainly along the
PIL. Then, at 08:41\,UT, a new system of loops centered at around $(x,y)\approx$ ($219$\arcsec, 
$-438$\arcsec), which connects the umbrae of opposite polarities, became visible for the first
time (see online movie). This coronal structure became more clear at 08:43\,UT and was best
visible at 08:45\,UT (see last two panels of Fig.\,\ref{a171_loops}). The indications of
the observed loops remained detectable at least until 08:51\,UT. Thus, their appearance is
co-temporal with scan 3 from the VTT. However, we cannot exclude that the loop system was present
in the upper atmosphere much earlier than discussed above and became temporarily visible
later when it was filled with plasma heated by the M-class flare. We note that this structure was
not visible in the AIA 1600\,\AA\ and 1700\,\AA\ images, which map a lower height in the atmosphere.
\begin{figure}[!t]
  \centering
  \includegraphics[width=\columnwidth]{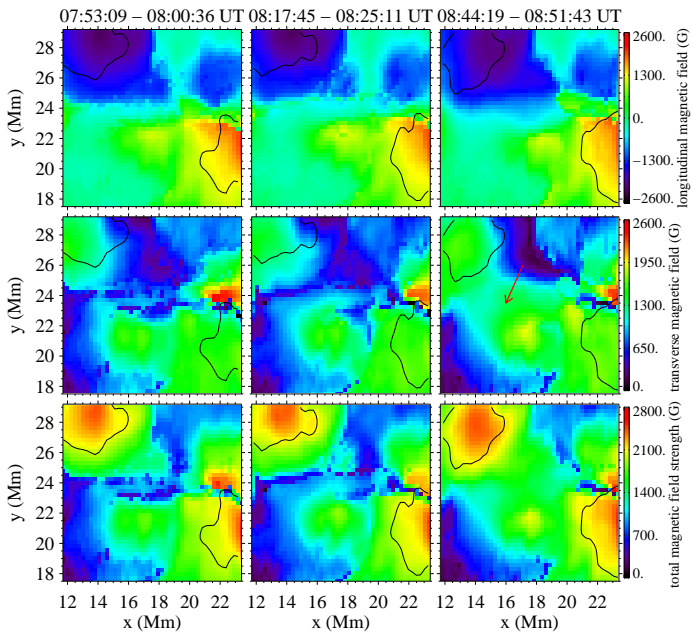}
     \caption{{\em From top to bottom:} longitudinal, transverse, and total magnetic 
              field strength inferred from the \ion{Fe}{i} 10783\,\AA\ inversions.
              Cutouts from scans 1, 2, and 3 showing internal area of the $\delta$-spot
              related to the observed changes of the magnetic field are displayed
              {\em from left to right}. The axes in $Mm$ are used in order to allow
              easier verification of determined parameters listed in the text (e.g.,
              values entering Equation \ref{eq1}). The red arrow marks an enhanced patch
              of transverse magnetic field detected in scan 3 (i.e., during the M-class
              flare). The black lines represent umbrae of the $\delta$-spot.}
     \label{B_Fe}
\end{figure}

The strongest flare was also associated with the eruption of a filament and coronal loop
system as observed in AIA, and a slow and faint CME (v $\approx$ 220 km\,s$^{-1}$) observed
by SOHO/LASCO (for more details see LASCO CME catalog: https://cdaw.gsfc.nasa.gov/).

\subsection{Magnetic field}
The SIR results provide the magnetic field strength inferred from the Stokes parameters.
The longitudinal and transverse components of the magnetic field are computed by multiplying
the magnetic field strength by $\cos\theta$ and $\sin\theta$, respectively, where $\theta$
is the line-of-sight (LOS) inclination of the magnetic field lines. We were limited to this
approach as we were not able to solve the azimuth ambiguity because of the complicated
magnetic structure of the observed $\delta$-spot.

\begin{figure}[!t]
  \centering
  \includegraphics[width=\columnwidth]{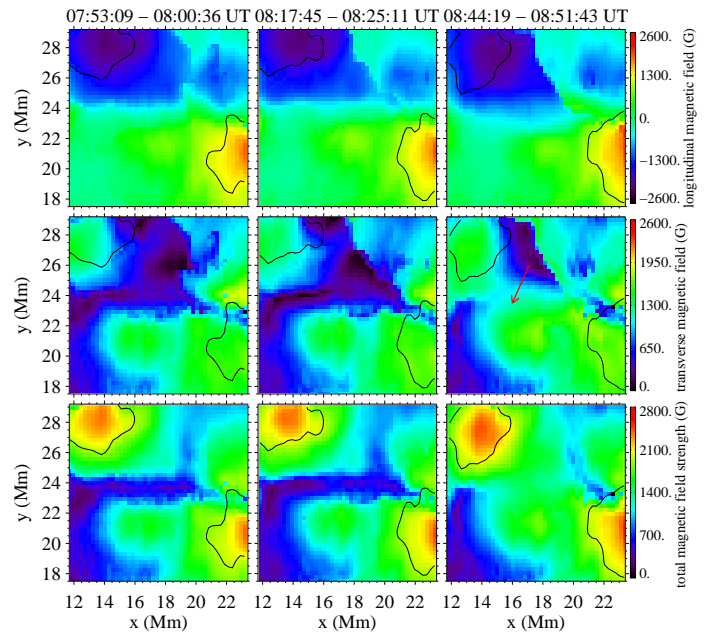}
    \caption{As in Fig.\,\ref{B_Fe} but for the inversions of \ion{Si}{i} 
             10786\,\AA\ spectral line.}
    \label{B_Si}
\end{figure}

The \ion{Si}{i} spectral line forms roughly 100--150\,km higher compared to \ion{Fe}{i}.
Both lines sample very low atmospheric heights
\citep{2008A&A...488.1085B}.                                           
Therefore, the overall inferred magnetic field strength is similar at both heights, but
as expected, it is slightly weaker in the \ion{Si}{i} line (see Figs.\,\ref{B_Fe} and \ref{B_Si}).

In all three scans, the longitudinal component of the magnetic field does not show any
significant changes. In general, only small and spatially very localized variations
(i.e., small increases as well as decreases) were detected. Thus, probably the only
noteworthy change is the appearance of the negative longitudinal field inside the 
northern part of PIL. This is seen as a slight downward deformation of the negative polarity
patch visible at coordinates around $(x,y)\approx$ ($15-20$\,Mm, $23-24$\,Mm) in the
top-rightmost panels (corresponding to scan 3) of Figs.\,\ref{B_Fe} and \ref{B_Si}.

\begin{figure}
   \centering{
    \includegraphics[width=\columnwidth]{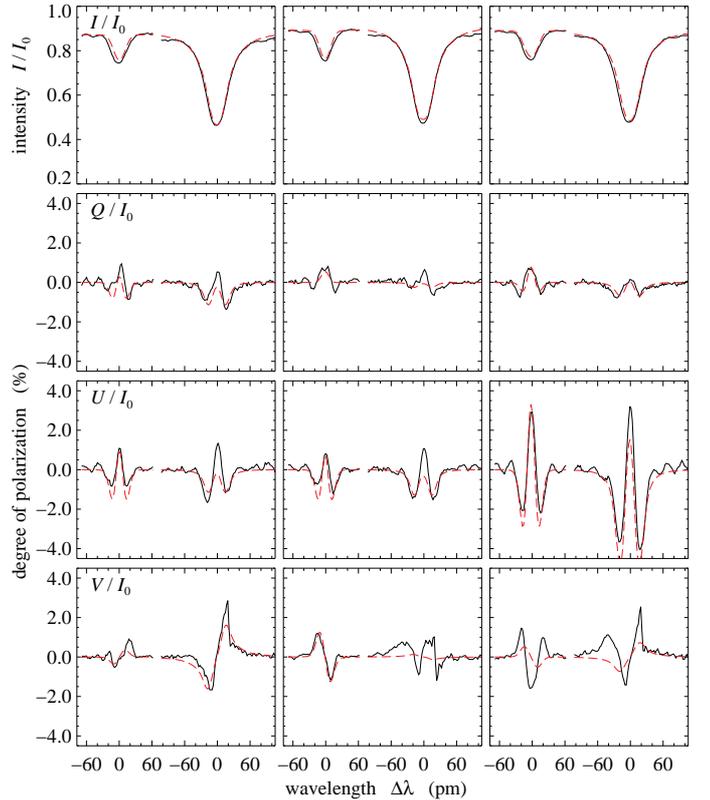}}
     \caption{Observed Stokes profiles for a selected location near the PIL displayed
              as solid black lines, and the results from the SIR-inversion overlaid as
              red dashed lines. In the left part of each panel, \ion{Fe}{i} 10783\,\AA\
              is shown, and in the right part, \ion{Si}{i} 10786\,\AA. The columns
              stand for the three scans, and from top to bottom, Stokes $I/I_{0}$,
              $Q/I_{0}$, $U/I_{0}$ and $V/I_{0}$ are shown.
              }
     \label{stokes_prof}
\end{figure}

However, a significant patch of the transverse component appeared at the PIL at around
$(x,y)\approx$ ($14-20$\,Mm, $20-25$\,Mm) (see the red arrow in Figs.\,\ref{B_Fe} and
\ref{B_Si}) in the last scan (after the flare). This patch bridges the PIL and connects
the umbrae within the $\delta$-spot. Fig.\,\ref{stokes_prof} shows the observed and
calculated Stokes profiles for one pixel next to the PIL where we encounter the increase
of the horizontal magnetic field during the flare. In general we see a reasonable agreement
of observed and inverted profiles. Therefore, a one-component inversion applied to data
is justified. Only for the few cases of anomalous but weak Stokes-$V$ profiles could
a two component inversion yield better fits. Nevertheless, there is an obvious increase
of the Stokes-$U$ amplitude after the flare.

An increase of approximately 550\,G with respect to scans 1 and 2 is visible in Figs.\,\ref{B_Fe}
and \ref{B_Si} in the area marked by the red arrow. The bottom panels show the total
magnetic field, which in turn also shows an increase in the above mentioned area.  

The newly occurred patch of increased transverse magnetic field is co-spatial and co-temporal
with the newly formed system of hot flare loops visible in the AIA 171\,\AA\ images, which
also connects umbrae within the observed spot (cf. Fig.\,\ref{a171_loops}). Thus, the natural
question is whether these two magnetic structures visible at very different temperatures
(and also heights) are related to each other or not.

An inspection of the HMI magnetograms between 07:45 and 09:00 UT (co-temporal to the VTT
observations) shows that both polarities were close together in the $\delta$-spot, only
separated by a thin PIL. As the M-class flare approached in time, the positive polarity
next to the PIL gradually vanished, that is, LOS magnetic flux slowly disappeared.
This visually produces a widening of the PIL. Most of the positive polarity
close to the border of the PIL disappeared once the flare happened.

\subsection{Doppler velocities}
The LOS velocities derived from the \ion{Fe}{i} and \ion{Si}{i} inversions are 
shown in Fig. \ref{tip_vel}. The same small cutout of the $\delta$-spot as in
Figs.\,\ref{B_Fe} and \ref{B_Si} is shown. The panels representing scan\,1 show an
extended redshifted area bridging both umbrae. Moreover, there is a patch of strong
redshifts (visible mainly in \ion{Fe}{i}) close to the umbra of positive polarity.
However, this patch disappears already in the second scan and only blueshifts are
seen along almost the whole PIL. These significant changes are probably caused by
the high flaring activity during the observations. On the other hand, blueshifts at
the PIL are clearly visible also in the panels corresponding to scan\,3. Thus, this
structure was not significantly affected by the M-class flare and surprisingly also
not by a newly formed patch of the transverse magnetic component. This is also supported
by HMI Dopplergrams (not shown here), which show predominantly blueshifted PIL during
the whole M-class flare.

\section{Discussion} \label{Discussion}
The $\delta$-spot was embedded in an active region that produced many flares over its lifetime 
(see Figs.\,\ref{fig_goes}, \ref{fig_fermi}, and \ref{a1600_activity}). All AIA wavelength
bands showed eruptive events during our ground-based observations. The HMI magnetograms
showed a widening of the PIL, especially after the eruption of an M-class flare, revealing
that the magnetic field topology was undergoing changes. Supporting evidence for this
can be found in the inferred magnetic field vector from the two low-lying photospheric
\ion{Fe}{i} 10783\,\AA\ and \ion{Si}{i} 10786\,\AA\ spectral lines. After the flare,
a patch of transverse magnetic field developed at the PIL (see Figs.\,\ref{B_Fe} and
\ref{B_Si}), which  increased the total magnetic field in the corresponding region as well.

This result fits well into recent predictions
\citep[e.g.,][]{2008ASPC..383..221H}                            
and observations
\citep[e.g.,][]{2012ApJ...745L..17W}                            
where the photospheric magnetic field tilts toward a more horizontal configuration
after a flare, that is, parallel to the surface. Flare-associated changes of the magnetic
field in observations and simulations were also studied by
\citet[][]{2011ApJ...727L..19L}                                 
who similarly reported an enhancement in the horizontal magnetic field near the
flaring PIL. They explain these changes as a natural consequence of the lift-off
of the pre-existing coronal flux rope, and the subsequent implosion of the magnetic
field with inward collapsing post-reconnection loops above the PIL. The topical M-class
flare was followed by the appearance of a new system of coronal loops connecting umbrae
within the $\delta$-spot as seen in the lower rightmost panel of Fig.\,\ref{context}
and in Fig.\,\ref{a171_loops}. They may correspond to the lifted-off flux rope
described in
\citet[][]{2011ApJ...727L..19L}.                                
A comprehensive study of flare-induced changes in the photospheric and chromospheric
magnetic field is presented in
\citet[][]{2017ApJ...834...26K}.                                
The author found that the changes of the photospheric magnetic field due to an X1 flare
are located near the PIL.
\begin{figure}[!t]
\centering
  \includegraphics[width=\columnwidth]{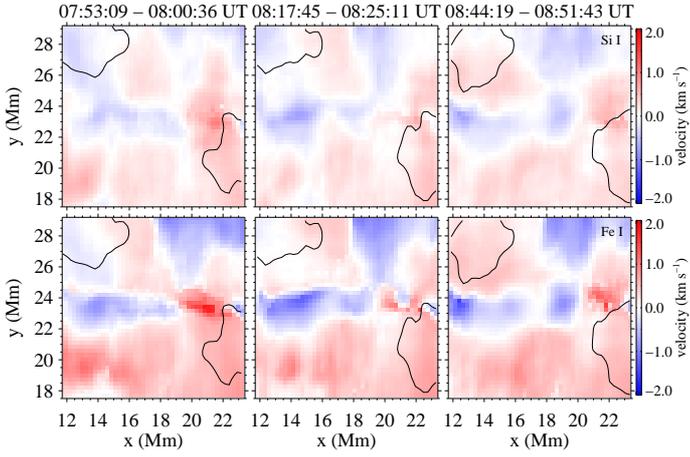}
     \caption{LOS velocities inferred using SIR inversions of
              \ion{Fe}{i} (bottom panels) and \ion{Si}{i} (top panels)              spectral lines. The velocities are clipped between $\pm 2$\,km\,s$^{-1}$.
              The black contours represent the umbrae of the $\delta$-spot.}
     \label{tip_vel}
\end{figure}

A detailed inspection of our results suggests that the changes of the photospheric magnetic
field closely coincide with the PIL and seem to follow its shape (Figs.\,\ref{context},
\ref{B_Fe}, and \ref{B_Si}). These results also suggests that the field topology suffered
the changes mostly in an elongated PIL-aligned area centered at $(x, y) \approx (15, 24)$\,Mm
with an approximate size of $5 \time 1$\,Mm (region marked by the red arrow in
Figs.\,\ref{B_Fe} and \ref{B_Si}), where the area-averaged changes of both the transverse
field $\left<\delta B_{TR}\right>$ and total magnetic field increased, on average, by approximately 550\,G. The changes of the longitudinal magnetic field $\left<\delta B_{LOS}\right>$ do not
reach more than 200\,G in the area close to or within the PIL.

\citet[][]{2012SoPh..277...59F}                                  
used the concept of momentum conservation and assessed the Lorentz-force changes due to
the release of energy associated with the rapid evolution of the coronal magnetic field.
They argue that a back-reaction following flares could push the photospheric magnetic
fields to become more horizontal near the flaring PIL. These authors estimated the changes
of the radial Lorentz force $\delta F_{r}$ (LF) exerted on the photosphere from the corona
to be
\begin{equation}
\label{eq1}
\delta F_{r} = \frac{1}{8\pi}\int_{A_{ph}}\left( \delta B_{LOS}^2 - \delta B_{TR}^2\right)dA \approx \frac{1}{8\pi}\left\{ \left<\delta B_{LOS}\right>^2 - \left<\delta B_{TR}\right>^2 \right\}A_{ph}
.\end{equation}

The surface integral involves the photospheric domain experiencing the magnetic field
changes. From its estimated area $A_{ph} \approx 5$\,Mm$^2$ and the values given before
follows $\delta F_{r} \approx 5.3\times10^{20}$\,dyn. This value is almost one order of
magnitude smaller than the downward LF change of $3.2\times10^{21}$\,dyn estimated in
\citet[][]{2013IAUS..294..561S}                                   
for the B4.2-class flare in AR NOAA 8027 on 1997 April 9, which was caused by a tiny
negative feature invading its surrounding positive polarity region. Our estimate
is also almost two orders of magnitude smaller than the LF change of $1.6\times10^{22}$\,dyn
reported in
\citet[][]{2010ApJ...716L.195W}                                   
for the 2002 July 26 M8.7 flare. They found similar values of the LF change for
several other X-class flares. For comparison,
\citet[][]{2016ApJ...820L..21X}                                   
estimated the LF peak value at approximately $1\times10^{22}$\,dyn, which was associated with
the X1.1 flare in the $\delta$-configuration region in the following part of AR 11890.

Unfortunately, in our case, the slit did not catch the highest emission of the flare. 
Therefore, we cannot estimate the changes in the magnetic field during the peak of the
flare. Such a study is reported by 
\citet[][]{2015ApJ...799L..25K,2015IAUS..305...73K}                
who showed a huge drop in the photospheric magnetic field during the maximum of the flare.

Remarkably, the LOS velocity maps in Fig.\,\ref{tip_vel} do not show any significant
changes of velocity patterns, which might be associated with the observed M-class
flare. The area coinciding with the magnetic field changes at $(x, y) \approx$ (15, 24)\,Mm
at the PIL is characterized by a persistent upflow of approximately $-1$\,kms$^{-1}$. This
behavior is confirmed by HMI Dopplergrams, which sample the region with much better
temporal resolution.

The quasi-invariability of the upflow could be due to very different optical depths,
where the velocity response function and the magnetic field response function have
their maxima. Another possibility to explain the ground-based observations would be
that the slow scanning simply missed the moment of occurrence of short-living/temporary
velocity changes induced by the flare. However, this scenario is excluded by SDO/HMI
measurements. On the other hand, we cannot exclude that the spatial resolution of our
observation was insufficient to resolve the probably very subtle flare-induced
velocity changes. 
\citet[][]{2015ApJ...799L..25K}                                    
reported on newly appearing upflows in response to an M3.2 flare. Although their
data were taken with the same instrument, the upflows were stronger next to the
PIL rather than along it. However, their flare was not related to a $\delta$-spot
and might consequently show different properties. Steady, persistent, and subsonic
motions were observed along the PIL of a $\delta$-spot also by 
\citet[][]{2014ApJ...789..162C}.                                    
They reported a stable pattern of upflows and downflows of $\pm$3\,km\,s$^{-1}$
along the PIL detected by ground-based high-resolution observations. The presence
of this pattern was confirmed by HMI measurements. Based on satellite data,
they estimated that the flows lasted for at least 15 hours. 

Finally, we recall that the increase of the transverse photospheric field is cospatial
and cotemporal with the new, hot flare loops visible in the AIA 171\,\AA\ channel (the
lower rightmost panel of Fig.\,\ref{context} and Fig.\,\ref{a171_loops}). Likely,
these are only two facets of magnetic field changes cascading across the flaring atmosphere,
yet we cannot confirm this hypothesis using our data. 
\section{Conclusions} \label{Conclusions}
We present rare ground-based spectro-polarimetric observations of the flaring
$\delta$-spot performed by the TIP-II instrument attached to the spectrograph of
the VTT. We found very stable magnetic configuration which remains almost unchanged
during a sequence of four C-class flares. However, after the strongest flaring event,
we discovered a significant increase of approximately 550\,G in the transverse magnetic field
in places connecting umbrae of opposite polarities. The observations taken by the SDO/AIA
instrument in the 171\,\AA\ channel revealed the formation of the new system of magnetic
loops which also connects umbrae within the observed $\delta$-spot. This suggests that
formation of a new system of loops reaching from photosphere up to corona was induced/triggered 
by the M-class flare. The Doppler velocities estimated from the spectro-polarimetric
scans as well as from measurements provided by the SDO/HMI instrument do not show any
significant changes in the plasma flows during formation of the aformentioned magnetic loops.
 
\begin{acknowledgements}
  This work was supported by the project VEGA 2/0004/16 and by the project of
  the \"{O}sterreichischer Austauschdienst (OeAD) and the Slovak Research and
  Development Agency (SRDA) under grant Nos. SK 01/2016 and SK-AT-2016-0002.
  A.M.V. gratefully acknowledges support from the Austrian Science Fund (FWF):
  P27292-N20. The authors deeply thank C. Denker for giving many comments on
  the manuscript to improve its quality. The authors thank an anonymous
  referee for constructive comments and valuable suggestions to improve this
  paper. This article was created by the realisation of the project ITMS No.
  26220120029, based on the supporting operational Research and development
  program financed from the European Regional Development Fund. The observations
  were taken within the SOLARNET Transnational Access and Service Programme.
  The Vacuum Tower Telescope in Tenerife is operated by the Kiepenheuer-Institut
  f\"{u}r Sonnenphysik (Germany) at the Spanish Observatorio del Teide of the
  Instituto de Astrof\'{i}sica de Canarias. SDO is a mission for NASA’s Living
  With a Star (LWS) Program. This research has made use of NASA’s Astrophysics
  Data System.
\end{acknowledgements}


\end{document}